\documentclass[12pt]{JHEP3}

\let\ifpdf\relax
\usepackage{ifpdf}
\usepackage{graphics}
\usepackage{graphicx}
\usepackage{amssymb}
\usepackage{amsmath}
\usepackage{slashed}
\usepackage{cite}
\usepackage{epsfig}
\usepackage{datetime}

\newcommand{\be}{\begin{equation}}
\newcommand{\ee}{\end{equation}}
\newcommand{\bal}{\begin{align}}
\newcommand{\eal}{\end{align}}
\newcommand{\bea}{\begin{eqnarray}}
\newcommand{\eea}{\end{eqnarray}}

\def\o{\omega}

\def\Tr{{\rm Tr}}

\def\ckn{C_{k/n}}
\def\bkn{\beta_{k/n}}

\def\2kn{\frac{2\pi k}{n}}

\def\wllk{W_{1}^{1(k)}}

\def\wzzk{W_{2}^{2(k)}}

\def\sun1{{\widehat{\rm SU}(N)_1}}
\def\un1{{{\rm U}(N)_1}}
\def\hatun1{{\widehat{\rm U}(N)_1}}

\title{R\'enyi Entropy for a $\bf 2d$ CFT with a gauge field: $\bf \sun1$ WZW theory on a branched torus}

\author{Howard J. Schnitzer  \\
{\small  Martin Fisher School of Physics, Brandeis University, \\ \ \ \ \ \ Waltham, MA 02454, USA\\
}

E-mail:  \email{schnitzr@brandeis.edu}}

\preprint{
 BRX-TH-6307}

\abstract{

The R\'enyi entropy for the $\sun1$ WZW model as described by $N$ free fermions coupled to a $U(1)$ constraint field is computed on an $n$-sheeted branched torus. The boundary condition of the harmonic component of the gauge field on the homology cycles of the genus $g$ Riemann surface is central to the final result. 

This calculation is complementary to that of arXiv:$1510.05993$, which presents the bose side of the bose-fermi equivalence. 
}

\begin{document}

\section{Introduction}
The treatment of entanglement entropy (EE) and R\'enyi entanglement entropy (RE) in the presence of gauge fields in still a developing open issue. The problem is that one cannot express the Hilbert space of the theory as a direct product 
${\cal H}\neq {\cal H}_A\otimes {\cal H}_B$, separating two regions $A$ and $B$ because a Gauss law constraint is not localized. Several proposals have been advanced usually evaluating the theory on a lattice, with suitable constraints on the boundary links of the lattice. A number of studies have pursued this approach 
\cite{Velytsky:2008rs,Buividovich:2008kq,Buividovich:2008gq,Nakagawa:2009jk,Klebanov:2011td,Eling:2013aqa,Agon:2013iva,Casini:2013rba,Donnelly:2011hn,Donnelly:2014gva,Casini:2014aia,Casini:2015dsg,Donnelly:2014fua,Donnelly:2015hxa,Huang:2014pfa,Ghosh:2015iwa,Aoki:2015bsa,Chen:2015kfa,Hung:2015fla,Radicevic:2014kqa,Radicevic:2015sza,Soni:2015yga,VanAcoleyen:2015ccp,Radicevic:2016tlt,Delcamp:2016eya}.

One strategy is to extend the Hilbert space ${\cal H}\to {\cal \overline{H}}$, so that for the extended space ${\cal \overline{H}}={\cal \overline{H}}_A\times{\cal\overline{H}}_B$
is realized, compute EE or RE on the extended space and project the final result to $\cal H$. To date this approach has only carried out on a lattice.

In this paper we discuss the RE for the $\sun1$ WZW model on a branched torus. This theory has the advantage that there are two presentations of the theory as a consequence of the bose-fermi equivalence. In the bose formulation there are no gauge fields, and the RE can be unambiguously calculated\cite{Schnitzer:2015ira,Naculich:1989ii}. In the fermi presentation
\bea 
\label{1.1}
\widehat{\rm SU}(N)_1=\frac{\hat{{\rm U}}(N)_1}{{\rm U}(1)}
\eea
there are $N$ free fermions constrained by a U$(1)$ gauge field. In this sense, the extended Hilbert space $\cal\overline{H}$ is that of $\un1$, and the final result for the RE is obtained when the constraint gauge field is considered. The consequences of the U$(1)$ gauge field are non-trivial as fermions in the presence of a constant gauge field is equivalent to free fermions with twisted boundary conditions\cite{Atick:1987kd},\cite{AlvarezGaume:1986es}.

In eq (\ref{1.1}) for genus $\geq 1$ one expands the gauge field in a Hodge decomposition. The exact and co-exact fields in the decomposition contribute to the RE in a straight forward way, essentially as scalar fields. The harmonic component of the gauge field contributes a fermi determinant with twisted boundary conditions, where the contributions of the harmonic component of the constraint gauge field removes any U$(1)$ charge circulating around a homology cycle.

This study of the RE for $\sun1$ on a branched torus examplifies the strategy of extending the Hilbert space ${\cal \overline{H}}={\cal \overline{H}}_A\times{\cal\overline{H}}_B$ and projecting the final result. The computation of this paper and of \cite{Schnitzer:2015ira} are heavily dependent of the methods developed in \cite{Naculich:1989ii}, and the bose-fermi equivalence demonstrated for $\sun1$ for arbitrary genus in that paper. The bose-fermi equivalence used here gives us confidence in the strategy employed in this paper. A challenge is to extend these ideas to other examples with continuum gauge fields, particularly when there is not a bose-fermi equivalence available to check the results.

Some papers which we found useful for our work are \cite{Chen:2015cna,
	Lokhande:2015zma,Calabrese:2009ez,Calabrese:2010he,Chen:2014ehg,Cardy:2014jwa,Chen:2014hta,Datta:2013hba,Azeyanagi:2007bj,Chen:2013kpa,Chen:2015uia,Chen:2015usa,Caputa:2015tua,Chen:2015kua,Chen:2013dxa,Headrick:2010zt,Calabrese:2004eu,Calabrese:2009qy,Hartman:2013mia,Cardy:2007mb,DeNobili:2015dla,Casini:2009sr,Casini:2008wt,Casini:2012ei,Liu:2015iia,Headrick:2015gba,Aoki:2016lma}

\section{The Action}
We are interested in the R\'enyi entropy for the WZW theory $\sun1$ on an $n-$branched torus, where 
\bea
\widehat{\rm SU}(N)_1=\frac{\hat{{\rm U}}(N)_1}{{\rm U}(1)}
\eea
is described by $N$ free fermions, constrained by a U$(1)$  gauge field. This approach is complementary to that of the bose formulation of the same theory \cite{Schnitzer:2015ira}, where a free boson field takes values on a  $N-$dimensional lattice. Here we concentrate on the fermi presentation of the theory, where $\hatun1$ spans a Hilbert space $\cal \overline{H}$ which is projected to $\sun1$, by the U$(1)$ gauge field.

We compute the R\'enyi entropy by means of the replica trick, where the R\'enyi entropy is
\bea
S_n=-\frac 1{n-1}\log \Tr \rho_A^n
\eea
In two-dimensional space-time, we consider the submanifold as that of a single interval on the torus. The two-dimensional theory is defined on a complex plane so that the R\'enyi entropy becomes
\bea
S_n=-\frac 1{n-1}\log \left(\frac{Z_n}{Z_1^n} \right) 
\eea
where $Z_n$ is the partition function for the $n-$sheeted Riemann surface which results from connecting the $n-$complex surfaces along the branch cuts; here the single interval. For a single interval we denote the surface as ${\cal R}_{n,1}$.

The Lagrangian density does not depend explicitly on the Riemann surface, so that the structure of ${\cal R}_{n,1}$ is implemented by appropriate boundary conditions. The partition function is  schematically of the form 
\bea\label{2.4}
Z_{{\cal R}_{n,1}}=\int [d\varphi] \exp\{-S_n[\varphi]\}
\eea
where $\varphi$ denotes all the various fields of the theory.
[Below we will be more explicit about the specific fields involved in our calculation.] We consider $n-$copies of the theory, with the partition function (\ref{2.4}) written as a path integral on the $n-$sheeted torus.
\bea\label{2.5}
Z_{{\cal R}_{n,1}}=\int_{\mathbb{C}_1}[d\varphi_1 \cdots d\varphi_n]\exp\{ \-\frac1{2\pi} \int_{\mathbb{C}}     (2i dz \wedge d\bar{z})[\mathcal{L}[\varphi_1]+\cdots +\mathcal{L}[\varphi_n]]  \}
\eea
where $\int_{\mathbb{C}_1}$ is the restricted path integral with appropriate boundary conditions. 

For $\hat{\rm U}(N)_1/$U$(1)$ the partition function is schematically as in (\ref{2.5}), where
\bea\label{2.6}
Z_{{\cal R}_{n,1}}=\int [d\chi] [d\tilde{\chi}][dA_\mu] [d\bar{A}_{\mu}] \exp(-I_{F}[\chi;A])_n
\eea
with the Lagrangian evaluated on ${\cal R}_{n,1}$. One copy of the Lagrangian as in (\ref{2.5}), (\ref{2.6}), is 
\bea \label{2.7}
I_F[\chi,A]=\frac1\pi \int d^2z [\bar{\chi}_a(z)(\bar{\partial}+i\bar{A})\chi^a(z)
+\tilde{\bar{\chi}}_a(\bar{z})(\partial+i A)\tilde{\chi}^{\bar{a}}(\bar{z})]
\eea
where $a=1$ to $N$,  $\chi^a$ and $\tilde{\chi}^{\bar{a}}$ are independent representations of U$(N)$ and the conjugate representations, respectively. The $A_i(z,\bar{z})$, $i=1,2$ are components of a U$(1)$ gauge field. In eqn. (\ref{2.7})
\bea
A=A_z=\frac12(A_1-iA_2) \nonumber \\
\bar{A}=A_{\bar{z}}=\frac12(A_1+iA_2)\,.
\eea
One can make the Hodege decomposition on a genus $g$ Riemann surface with the harmonic component $A^0$ 
so that 
\bea
A_\mu dx^\mu =Adz +\bar{A}d\bar{z}+A^{0}
\eea
where  
\bea
A^{0}=2\pi \sum_{a=1}^g [u_a \alpha_a-v_a \beta_a]
\eea
and where $u_a$ and $v_a$ are real constants, with the homology basis $(a_a,b_a)$ of one-forms, such that 
\bea\label{2.11}
\int_{a_a} \alpha_b=\delta_{ab}\, ; \qquad && \int_{b_a} \alpha_b =0 \nonumber \\
\int_{a_a} \beta_b=0 \, ; \qquad &&\int_{b_a} \beta_b =\delta_{ab}
\eea
Change variables on each copy of the Lagrangian density, i.e. 
\bea \label{2.12}
A=i(\partial \bar{h})\bar{h}^{-1} \nonumber \\
\bar{A}=i(\bar{\partial}h)h^{-1} 
\eea
where $h$ and $\bar{h}$ are diagonal $N\times N$ matrices belonging to U$(1)_c$ (i.e. complexified U$(1)$). The reality condition $A^\dagger=\bar{A}$ imposes $\bar{h}^{\dagger}=h^{-1}$, where the matrices $h$ and $\bar{h}$ are 
\bea \label{2.13}
h=e^{-\rho-i\lambda} \mathbb{I} \nonumber \\
\bar{h}=e^{\rho-i\lambda} \mathbb{I}
\eea
and where $\mathbb{I}$ is the $N\times N$ unit matrix. Thus for each copy of the Lagrangian 
\bea \label{2.14}
A=\partial \lambda+i\partial \rho \nonumber \\
\bar{A}=\bar{\partial} \lambda-i\bar{\partial} \rho
\eea
Upon changing variables, (\ref{2.12}, \ref{2.13}, \ref{2.14}), eqn. (\ref{2.7}) becomes schematically 
\bea\label{2.15}
Z_{\mathcal{R}_{n,1}}=Z_{gh} \int [d\chi][d\bar{\chi}] [dh][d\bar{h}][dA^0] \exp(-I_F[\chi, A])_n
 \eea
where $Z_{gh}$ denotes the Jacobian induced by the change of variables  (\ref{2.12}, \ref{2.13}, \ref{2.14}). 
The coupling of the fermions to $\lambda$ and $\rho$ can be rotated away by gauge and chiral transformations, at the price of a chiral anomaly $I_B[e^{2\rho} \mathbb{I}]$, but the coupling of the fermions to the harmonic component $A^0$ remains.  Choose the gauge $\lambda=0$, which involves no non-trivial Jacobian, allowing one to drop the $\lambda$ integration. The gauge coupling to the fermions becomes 
\bea\label{2.16}
\chi=h\psi ; \qquad\tilde{\chi}=\tilde{\psi} h^{-1}\nonumber \\
\bar{\chi}=\bar{h}\bar{\psi} ; \qquad  \tilde{\bar{\chi}}=\tilde{\bar{\psi}} \bar{h}^{-1}
\eea 
The measure is not invariant under the chiral transformation, but picks up a factor of $\exp(I_B[h^{-1}, h])$, where $I_B[g]$ is the WZW action 
\bea
\label{2.17}
I_{B}[g]=-\frac{1}{8\pi}\int d^2z \Tr[g^{-1}(\partial_i g)g^{-1}(\partial^i g)]
\eea
for $g\in$ U$(1)_c$. Since this group is abelian, there is no topological terms in (\ref{2.17}).

At this stage (\ref{2.15}) becomes
\bea\label{2.18}
Z_{\mathcal{R}_{n,1}}=Z_{gh} \int [d\psi][d\bar{\psi}] [dh][d\bar{h}][dA^0] \exp(-I_F[\psi, A^0])_n\exp(-I_B[h^{-1}, h])_n
\eea
Equation (2.18) is not quite correct, even as a formal expression, since the integration is over infinite volume, which can divided out. One arrives at 
\bea\label{2.19}
Z_{\mathcal{R}_{n,1}}=Z_{gh} \int [d\psi][d\bar{\psi}] [d\rho][dA^0] \exp(-I_F[\psi, A^0])_n\exp(-\frac{N}{2\pi}\int d^2 z(\partial_\mu\rho)(\partial^\mu \rho))_n \nonumber \\
\eea
One can change the normalization of $\rho$, as follows
\bea
\rho(z,\bar{z})=\frac 1{\sqrt{N}}\Phi (z,\bar{z}) 
\eea
The free boson action in (\ref{2.19}), now will have the normalization
\bea\label{2.21}
I_{B}[\Phi]&=&-I_B [e^{2\rho}]\nonumber \\
&=& \frac 1{2\pi} \int d^2z (\partial_i \Phi) (\partial^i \Phi)
\eea
The equations of motion imply
\bea\label{2.22}
\Phi(z,\bar{z})=\frac 12 [\phi(z)+\bar{\phi}(\bar{z})]
\eea
The final expression becomes
\bea\label{2.23}
Z_{\mathcal{R}_{n,1}}=Z_{gh} \int [d\psi][d\bar{\psi}] [d \Phi][dA^0] \exp(-I_F[\psi, A^0])_n\exp(-I_B[\Phi])_n \nonumber \\
\eea
where the ghost modes describe a CFT with central charge $c_{gh}=-2$. Equations (\ref{2.21}, \ref{2.22}) imply the genus zero two-point functions
\bea
\langle \phi(z) \phi(\omega) \rangle &=&-\log(z-\omega) \nonumber \\
\langle \bar{\phi}(\bar{z}) \bar{\phi}(\bar{\omega}) \rangle &=&-\log(\bar{z}-\bar{\omega}) 
\eea
To reiterate, the fermi and bose actions in (2.19) are given by $n$-copies, as schematically in (\ref{2.5})

\section{The Partition Function}
\subsection{Monodromies}
The fermions and bosons in (\ref{2.18}) obey different periodicity conditions in $j$, where $j$ labels the $j^{\rm{th}}$ Riemann sheet, $j=1$ to $n$. Generically, on circling a branch-point
\bea
\phi_j(ze^{2\pi i})&=&\phi_{j+1}(z)\quad {\rm for\quad bosons \quad and} \\
\psi_j(ze^{2\pi i})&=&\psi_{j+1}(z)\quad {\rm for\quad fermions } 
\eea
with
\bea
\phi_{j+n}=\phi_j
\eea
and 
\bea
\psi_{j+n}=(-1)^{n-1} \psi_j
\eea
The U$(N)$ free fermi fields satisfy
\bea
\psi^a_j(ze^{2\pi i})&=&\psi^a_{j+1}(z) \\
\bar{\psi}_{j,a}(ze^{2\pi i})&=&\bar{\psi}_{j+1,a}(z) \\
\tilde{\psi}^{\bar{a}}_j(\bar{z}e^{-2\pi i})&=&\tilde{\psi}^{\bar{a}}_{j+1}(\bar{z}) \\
\tilde{\bar{\psi}}_{j,\bar{a}}(\bar{z}e^{-2\pi i})&=&\tilde{\bar{\psi}}_{j+1,\bar{a}}(\bar{z}) 
\eea
on circling a branch point.\\
 The harmonic component of the U$(1)$ gauge field also satisfies the boson monodromies
 \bea
 A^0_j(ze^{2\pi i})=A^0_{j+1}(z) \\
 \bar{A}^0_j(\bar{z}e^{-2\pi i})=\bar{A}^0_{j+1}(\bar{z})
 \eea
 It is convenient to diagonalize the fields with respect to the monodromies ($k=1$ to $n$). Generically for the bosons
 \bea\label{3.11}
 \hat{\phi}_k(z)=\sum_{j=1}^n \phi_j(z) (\theta_j)^k
 \eea
 where
\bea
(\theta_j)^k=e^{2\pi i jk/n}
\eea
so that 
\bea
 \phi_j(z) =\sum_{k=1}^n (\theta^{-1}_j)^k\hat{\phi}_k(z)
\eea
where
\bea
(\theta^{-1}_j)^k=\frac 1n e^{-2\pi i jk/n}
\eea
Similarly diagonalizing the various fermions with respect to the monodromy\footnote{We use the monodromy of\cite{Giveon:2015cgs} as convenient. See also C. P Herzog and Y. Wang \cite{Herzog:2016ohd}}gives
\bea
 \hat{\psi}_k(z)&=& \sum_{j=1}^n\psi_j(z) \exp \frac{2\pi i j }{n}[k-\frac 12 (n-1)] \nonumber \\
 &=&\sum_{j=1}^n \psi_j(z) (\tilde{\theta}_j)^k
\eea
so that
\bea\label{3.16}
(\tilde{\theta}^{-1}_j)^k=\frac1{n}\exp -\frac{2\pi i j }{n}[k-\frac 12 (n-1)]
\eea

\subsection{The fermion action}
From (\ref{2.7}), (\ref{2.16}) and (\ref{2.18}) the fermion action is 
\bea\label{3.17}
I_F[\psi,A^0]_n
=\left\{  \frac 1\pi \int d^2z[\bar{\psi}_a(z)(\bar{\partial} + i A^0) \psi^a(z)+\tilde{\bar{\psi}}_{\bar{a}}(\bar{z})(\partial + i A^0)  \tilde{\psi}^{\bar{a}}(\bar{z}) ]   \right\}_n
\eea
Diagonalize the fields as in (\ref{3.11})-(\ref{3.16}), and then to simplify the notation, omit the hat in the fields. That is, write $\bar{\psi}_{a,k}(z)$ instead of $\hat{\psi}_{a,k}(z)$, etc. Then in terms of the diagonal fields, the Lagrangian density appropiate to (\ref{3.17}) becomes
\bea\label{3.18}
\frac 1n \left\{  \sum_{k=1}^n  \bar{\psi}_{a,k}(z) [\bar{\partial} +i\sum_{j=1}^n \bar{A}_j(z)(\frac 1n e^{-\frac{2\pi i}{n} (j-k)}) ]\psi^a_k(z) + {\rm h.c.}\right\}
\eea
The functional integral over the $N$ Dirac fermion coupled to the constant (in $z$) gauge field is equivalent to that of a free fermion with twisted boundary conditions. []

On a compact genus $g$ Riemann surface without branch-cuts, one can choose a canonical homology basis (\ref{2.11}). An alternate basis of one-forms are the holomorphic and anti-holomorphic differentials $\o_a$ and $\bar{\o}_a$ on the surface, whose integrals around the homology cycles in this basis are 
\bea\label{3.19}
\int_{a_a}\o_b=\delta_{a b } \qquad \int_{b_a}\o_b=\Omega_{a b } \nonumber \\
\int_{a_a}\bar{\o}_b=\delta_{a b } \qquad \int_{b_a}\bar{\o}_b=\bar{\Omega}_{a b }
\eea 
where $a=1$ to $g$, and $\Omega_{a b }$ is the $g\times g$ period matrix of the surface.\\
For a Riemann surface with branch-cuts, an appropriate basis is that of cut-differentials\cite{Atick:1987kd}. For the branched-torus we are considering, the appropriate basis are the doubly-periodic cut-differentials $\o^k_1(z)$ and $\o^k_2(z)$; $k=1$ to $n$ and their integrals over the intervals $0\leq z\leq 1$ and $0\leq z \leq \tau$ [Appendix A of \cite{Schnitzer:2015ira}]

The fermi action in the presence of a constant gauge field is equivalent to that of free fermions with twisted boundary conditions, where the result is expressed in terms of the period matrix of the surface \cite{AlvarezGaume:1986es}. Thus our task is to identify the period matrix for the surface, described by the basis of cut-differentials. To do so, one expands the gauge field in (\ref{3.18}) in this basis. Note the normalization 
\bea\label{3.20}
\frac{1}{\wllk}\int_0^1 dz \, \o_1^k&=&1\\\label{3.21}
\frac{1}{\wllk}\int_0^{\tau} dz \, \o_1^k&=&-\left[\frac{\wzzk}{\wllk} \right ]
\eea
for the torus (at each k) $0\leq \sigma_1\leq 1$ and $0\leq \sigma_2 \leq \tau$. \\

Therefore from (\ref{3.18}) the period matrix in this basis is
\bea\label{3.22}
\Omega_{rs}^{(k)}=\frac{16 \pi }{n} \left[\frac{\wzzk}{\wllk} \right ](e^{-2\pi i k/n})^{r-s}
\eea
where the overall normalization is chosen to agree with that of equation (4.19) in the bose formulation \cite{Schnitzer:2015ira}. Proceeding \footnote{We omit repeating these readily available details in the interest of clarity of the presentation } as in equation (6.7) of \cite{AlvarezGaume:1986es} and then (4.26)-(4.36) and (4.65) of \cite{Naculich:1989ii}, we find 
\bea\label{3.23}
Z_{fermi}=Z_{hom} \exp -\sum_{k=1}^n
S^{(k)}
\eea
where 
\bea\label{3.24}
   \bkn=\left | \frac{\wzzk}{\wllk} \right |
\eea   
   and 
\bea
\label{3.25}
S^{(k)}=\frac{16\pi}{n} \sum_{r,s=0}^{n-1}\sum_{a=1}^{N-1}\{ \bkn\, m_r^a (\ckn)_{rs}\,m_{s}^a \nonumber \\ \qquad \qquad + (\bkn)^{-1}\, n_r^a (\ckn)_{rs}\, n_{s}^a\}
\eea
with $m_r^a$ and $n_r^a \in \mathbb Z$. After summation over $k$, the phase in (\ref{3.17}) reduces to
 \bea\label{3.26}
(\ckn)_{rs}=\cos[\2kn(r-s)] 
\eea
In (\ref{3.23}) there is also a homogeneous contribution to the fermi action.

One performs a Poisson summation on (\ref{3.23})-(\ref{3.25}) with the result 
 \bea\label{3.27}
Z_{fermi}=Z_{hom} \prod_{k=1}^n (\bkn)^{1/2}\exp -\sum_{k=1}^n
S^{(k)}
\eea
where now 
\bea
\label{3.28}
S^{(k)}=\frac{16\pi}{n} \sum_{r,s=0}^{n-1}\sum_{a=1}^{N-1}(\bkn) \{  m_r^a (\ckn)_{rs}\,m_{s}^a \nonumber \\ \qquad \qquad + n_r^a (\ckn)_{rs}\, n_{s}^a\}
\eea
Finally from a calculation analogous to Sec 4.2 of \cite{Schnitzer:2015ira}, the homogeneous contribution is \footnote{There is also an overall normalization in the final partition function which one fixes to agree  with the bose-fermi equivalence}
\bea\label{3.29}
Z_{hom}=\frac{1}{|\eta(\tau)|^{2Nn}}\prod_{k=0}^{n-1}\left| \frac{\theta'_1(0|\tau)}{\theta_1(z_1-z_2|\tau)} \right|^{2\Delta_\psi}
\eea
where 
\bea
\Delta_\psi=\frac N2 (k/n)^2
\eea
\subsection{The remaining action}
From (\ref{2.19})-(\ref{2.21}) there remains 
\bea\label{3.31}
Z_{gh} Z_{\Phi}=Z_{gh} \int [d\Phi] \exp(-\hat{I}_B[\Phi])_n
\eea
with $\hat{I}_B$ given by (\ref{2.21}) and (\ref{2.22}). A calculation analogous to that of equation (4.30) of \cite{Schnitzer:2015ira}  gives
\bea
 Z_{\Phi}=\frac{1}{|\eta(\tau)|^{2n}}\prod_{k=0}^{n-1}\left\{   \frac{1}{|\wllk \wzzk |^{1/2}}  \left| \frac{\theta'_1(0|\tau)}{\theta_1(z_1-z_2|\tau)} \right|^{2\Delta_{\Phi}} \right\}
\eea
where 
\bea
\Delta_{\Phi}=\frac12 (k/n)^2
\eea
and 
\bea\label{3.34}
Z_{gh}=|\eta(\tau)|^{4n}\prod_{k=0}^{n-1}\left| \frac{\theta'_1(0|\tau)}{\theta_1(z_1-z_2|\tau)} \right|^{2\Delta_{gh}}
\eea
where 
\bea
\Delta_{gh}=-(k/n)^2
\eea

\subsection{The complete Action}
The complete action is 
\bea\label{3.36}
Z&=&Z_{fermi}Z_{gh}Z_{\Phi} \nonumber \\
&=&\frac{1}{|\eta(\tau)|^{2(N-1)n}}\prod_{k=0}^{n-1}\left\{   \frac{1}{|\wllk |}  \left| \frac{\theta'_1(0|\tau)}{\theta_1(z_1-z_2|\tau)} \right|^{2\Delta_{k/n}} \right\}  |\Theta(0|i\Omega)|^2
\eea
where 
\bea
\Delta_{k/n}=\frac{N-1}{2}(k/n)^2
\eea
and 
\bea
\Theta(0|i\Omega)=\sum_{\vec{m}\in \mathbb{Z}}\exp[-\pi \vec{m}_r\cdot C_{rs}\cdot\vec{m}_s]
\eea
The final result (\ref{3.36}) is normalized to be  identical to that of (4.32) of the bose formulation \cite{Schnitzer:2015ira}.

\section{Conclusions}
The existence of the bose-fermi equivalence for the $\sun1$ WZW model on the branched torus provides confidence in our treatment of the constraint $U(1)$ gauge field. However, since the RE depends on a Riemann-Siegel theta function, it is not possible to take the $n\to 1$ limit  to compute the EE. Nevertheless an important aspect of the calculation is the expansion of the $U(1)$ gauge field in a Hodge decomposition with the harmonic component essential to the final result, which ensures that there are no $U(1)$ charges circulating around the homology cycles. It is the contribution of this term that cannot be continued to $n\to 1$. On the other hand the closed and co-closed components lead to contributions analogous to that of conventional scalar field theories, so that the continuation to $n\to 1$ for these terms is not an issue, and that an extended Hilbert space is not needed here.

It is difficult to compare our results directly with that of lattice calculations where the set-up of an extended Hilbert space ${\cal \overline{H}}={\cal \overline{H}}_A\times{\cal\overline{H}}_B$ is also employed. Typically the lattice formulation provides constraints on the normal component of an electric field on the boundary of $A$, as well as a scalar field living on the boundary of $A$. Here the gauge field is likely analogous to a sector with E$_\perp = 0$ on the boundary. However note that both the classical and quantum contributions are necessary to obtain agreement with the bose formulation, while only the quantum piece is distillable \cite{Soni:2015yga},\cite{VanAcoleyen:2015ccp}.

There are some generalizations of the $\sun1$ WZW theory that can be considered. One can add a kinetic energy term or other terms depending on  the $U(1)$ gauge field, but there will no longer be a bose-fermi equivalence to check the results. One can also generalize the constraint to a non-Abelian gauge field so that $$ \widehat{\rm SU}(N)_K=\frac{\hat{{\rm U}}(NK)_1}{{\rm U}(K)}$$ can be considered.
However a comprehensive overall strategy for continuum gauge fields is still needed. 

For another application of fermi-bose equivalence to entanglements issues see \cite{Headrick:2012fk}, and for other aspects of bose-fermi equivalence see\cite{Redlich:1986rp,Schnitzer:1986fi}

\section*{Acknowledgements}


H. J. S. wishes to thank  Cesar Ag\'on and Isaac Cohen-Abbo for their invaluable assistance in preparing this paper.
The research of H. J. S. is supported in part by DOE by grant  DE-SC0009987. 

This paper represents a contribution to the celebration of the author's $82^{nd}$ birthday.

\newpage
\bibliographystyle{utphys}

\bibliography{emirefs}

\end{document}